\documentclass{elsart}
\usepackage{graphicx}

\begin{document}
\runauthor{Jensen}
\begin{frontmatter}
\title{Enumerations of plane meanders}
\author{Iwan Jensen}
\address{Department of Mathematics \& Statistics, 
The University of Melbourne,\\
Parkville, Victoria 3052, Australia \\
e-mail: i.jensen@ms.unimelb.edu.au}

\maketitle
\begin{abstract}
A closed plane meander of order $n$ is a closed self-avoiding loop 
intersecting an  infinite line $2n$ times. Meanders are considered 
distinct up to any smooth deformation leaving the line fixed. We have 
developed an improved algorithm, based on transfer matrix methods, for the 
enumeration  of plane meanders. This allows us to calculate the number of 
closed meanders up to $n=24$. The algorithm is easily modified to enumerate 
various systems of closed meanders, semi-meanders or open meanders.
\end{abstract}
\end{frontmatter}

\section{Introduction}

Meanders form a set of combinatorial problems concerned with the enumeration 
of self-avoiding loops crossing a line through a given number of points. 
Meanders are considered distinct up to any smooth deformation leaving the 
line fixed. This problem seems to date back at least to the work of 
Poincar\'e on differential geometry.  More  recently it has 
been related to enumerations of ovals in planar algebraic curves \cite{Arnold}
and the classification of 3-manifolds \cite{KS}. During the last decade or so 
it has received considerable attention in other areas of science. In computer 
science meanders are related to the sorting of Jordan sequences \cite{HMRT}. 
In physics meanders are relevant to the study of compact foldings of polymers 
\cite{FGG1,FGG2}, properties of the Temperley-Lieb algebra 
\cite{FGG3,Francesco}, and defects in liquid
crystals and $2+1$ dimensional gravity \cite{Kholodenko}.

The difficulty in the enumeration of most interesting combinatorial problems
is that, computationally, they are of exponential complexity. Initial efforts 
at computer enumeration of meanders have been based on direct counting. 
Lando and Zvonkin \cite{LZ} studied closed meanders, open meanders
and  systems of closed meanders, while Di Francesco {\em et al.} 
\cite{FGG2} studied semi-meanders. In this paper we describe a new
and improved algorithm, based on transfer matrix methods, for the 
enumeration of closed plane meanders. While the algorithm still has
exponential complexity, the growth in computer time is much slower
than that experienced with direct counting, and consequently the
calculation can be carried much further. The algorithm is easily modified to 
enumerate systems of closed meanders, semi-meanders or open meanders.

\section{Definitions of meanders \label{sec:def}}

\begin{figure}[h]
\begin{center}
\includegraphics[scale=0.8]{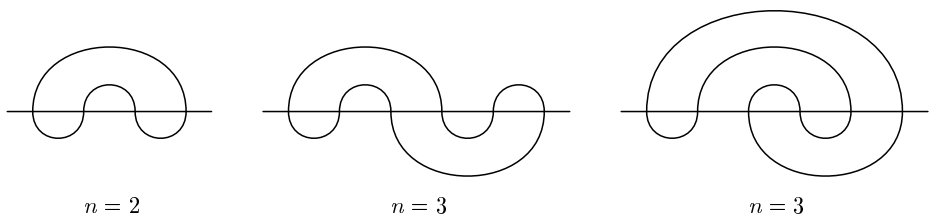}
\caption{\label{fig:meanclose} A few examples of closed meanders of
order 2 and 3, respectively. }
\end{center}
\end{figure}

A {\em closed meander} of order $n$ is a closed self-avoiding 
curve crossing an infinite line $2n$ times. Fig.~\ref{fig:meanclose} shows 
some meanders. The meandric number $M_n$
is simply the number of such meanders distinct up to smooth
transformations. We define the generating function

\begin{equation}\label{eq:meangen}
M(x) = \sum_{n=1}^{\infty} M_n x^n.
\end{equation}

We can extend the definition to {\em systems of closed meanders},
by allowing configurations with disconnected closed meanders.
The meandric numbers $M_n^{(k)}$ are the number of meanders with
$2n$ crossings and $k$ components.
An {\em open meander} of order $n$ is a self-avoiding curve running from 
west to east while crossing an infinite line $n$ times. The number
of such curves is $m_n$. It should be noted
\cite{LZ} that $M_n = m_{2n-1}$.
Finally, we could consider a semi-infinite line and allow the curve to wind
around the end-point of the line. A {\em semi-meander} of order $n$ is a 
closed self-avoiding loop crossing the semi-infinite line $2n$ times.  
The number of semi-meanders of order $n$ is denoted by $\overline{M}_n$.
In this
case a further interesting generalisation is to study the number
of semi-meanders $\overline{M}_n(w)$, which wind around the  end-point of 
the line $w$ times.

\section{Enumeration of meanders \label{sec:enum}}

\begin{figure}[h]
\begin{center}
\includegraphics[scale=0.9]{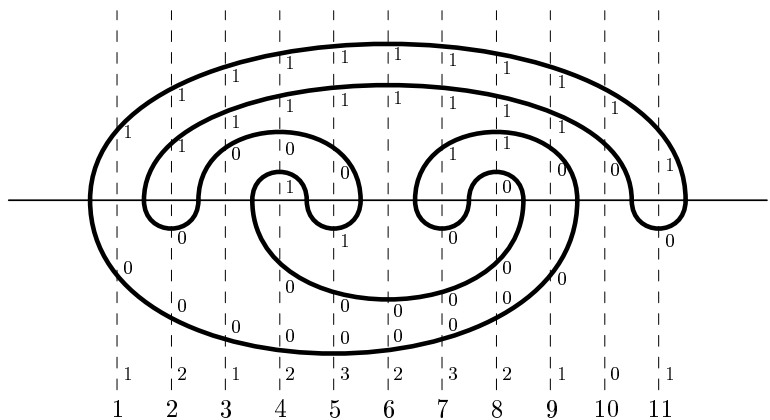}
\caption{\label{fig:transfer}
Positions of the boundarys (dashed lines) during the transfer matrix
calculation.  Numbers along the boundarys give the encoding of the 
loop structure in the partially completed meander to the left of 
the boundary.}
\end{center}
\end{figure}

The method used to enumerate meanders is similar to the transfer matrix
technique  devised by Enting \cite{Ent} in his pioneering work on the 
enumeration of self-avoiding polygons. The first terms in the series for 
the meander generating function can be calculated using transfer matrix 
techniques.  This involves drawing a boundary  
perpendicular to the infinite line. Meanders are enumerated by successive
moves of the boundary, so that one crossing at a time is added to the
meanders as illustrated in Fig.~\ref{fig:transfer}.
At each position of the boundary we have a configuration of loop-ends 
closed to the left, and for each configuration we count all the possible 
meanders that could give rise to that particular configuration of loop-ends.
Since the curve making up a meander is self-avoiding each configuration can 
be represented by an ordered set of edge states $\{x_i\}=0$ (1) indicates
the lower (upper) part of loop closed to the left of the boundary.
In addition we need to know where the infinite line is situated within the
loop-ends. This can be done simply by specifying how many loop-ends
lie beneath the infinite line. Configurations are read from the bottom to 
the top. As an example we note that the configuration along the boundary 
of the meander in Fig.~\ref{fig:transfer} at position 4 is $\{2;001011\}$.

We start with the configuration \{1;01\} with a count of 1, that is one 
loop crossing the infinite line. Next we move the boundary one step 
ahead and add a new crossing. So we either put in an additional loop or 
we take an existing loop-end immediately above or below the infinite 
line and drag it across the line. The first possibility is illustrated 
in Fig.~\ref{fig:transfer} in going to position 2 where we get the 
configuration \{2;0011\}. Additional loops are also inserted while going 
to positions 4 and 7. As we cross the infinite line with an existing 
loop-end we may be allowed to connect it to the loop-end on the other side. 
In going to position 6 we connect a `1' below the line to a `0' above the 
line and no further processing is required. In going to position 8 or 9  a 
`0' below the line is connected to a `0' above the line. This requires 
further processing because in connecting two lower loop-ends an 
upper loop-end elsewhere in the old configuration becomes a lower 
loop-end in the new configuration.
In going to position 8 we see that the configuration
\{2;000111\} before the step forward becomes the configuration
\{1;0011\} after the step. That is the upper end of the third loop
before the step becomes the lower end of the second loop after the step.  
We refer the reader to \cite{Ent} for the detailed  rules for 
relabeling of configurations.
Finally, note that connecting a `0' below the line to a
`1' above the line results in a closed loop. So this is only allowed if 
there are no other loops cut by the boundary and the result is a 
valid closed meander. 
As we move along and generate a new `target' configuration its count is 
calculated by adding up the count for the various `source' configurations 
which could generate that target. For example the `target' \{2;0011\} could 
be generated from the `sources'  \{1;01\}, \{1;0011\}, \{3;0011\} and
\{3;001011\}, by, respectively,  putting in an additional loop, moving a 
loop-end below the line, moving a loop-end above the line and connecting
two loop-ends across the line. 

The number of configurations, which need be generated in a calculation
of $M_{n}$, is restricted by the fact that at each step
we change the number of loop-ends above and/or below the infinite
line by at most one. So if we have already taken $k$ steps
then there can be at most $2n-k$ loop-ends above or below the line.
Any configurations violating this criterion can be discarded. Furthermore
we can reduce the number of distinct configurations by a factor of two 
by using the symmetry with respect to reflection in the infinite line.

%\subsection{Generalisations to other meander problems}

As we noted above  connecting a `0' below the line to a `1' above the line 
results in a closed loop. Failure to observe the restriction on this closure
would result in graphs with disconnected components, either one closed
meander over another or one closed meander within another. Obviously these 
are just the types of graphs required in order to enumerate systems of
closed meanders. So by noting that each such closure adds one more
component it is straightforward to generalise the algorithm to enumerate 
{\em systems of closed meanders}. 
{\em Open meanders} are a little more complicated. Suffice to say at this
stage that the main part of the necessary generalisations consists in 
adding an extra piece of information. We have to add a free end and specify 
its position within the configuration of loop-ends.
In order to enumerate {\em semi-meanders} all we just
change the initial configuration, and start in a position 
just before the first crossing of the semi-infinite line with $w$ loops
nested within one another. By running the algorithm for each
$w$ from 0 to $n$ we count semi-meanders with up to $n$
crossings.

%\subsection{Computational complexity \label{sec:complex}}

Using the new algorithm we have calculated $M_n$ up to $n=24$ as compared to 
the previous best of $n=17$ obtained by V. R. Pratt \cite{Sloane}. To fully 
appreciate the advance it should be noted that the computational 
complexity grows exponentially, that is the time required 
to obtain $n$ term grows asymptotically as $\lambda ^n$. For direct
enumerations time is simply proportional to $M_n$ and thus
$\lambda = \lim_{n \rightarrow \infty} M_{n+1}/M_n \approx 12.26 $.
The transfer matrix method employed in this paper is far more 
efficient and the numerical evidence suggests that the computational
complexity is such that $\lambda \approx 2.5$. 
Another way of gauging the improved efficiency is to note that
the calculations for semi-meanders carried out in \cite{FGG2} 
were ``done on the parallel Cray-T3D (128 processors) of the
CEA-Grenoble, with approximately 7500 hours $\times$ processors.''
Or in total about 100 years of CPU time. The equivalent calculations
can be done with the transfer matrix algorithm in about 15 minutes
on a single processor DEC-Alpha workstation!
The price we have to pay is that unlike for direct enumeration memory use 
grows exponentially with growth factor  $\lambda$. 

\section{Results and analysis \label{sec:results}}

\begin{table}[h]
\caption{\label{tab:closed} The number, $M_n$, of connected closed 
meanders with $2n$ crossings.}
\begin{center}
\begin{tabular}{rrrrrr} \hline \hline
$n$ & \multicolumn{1}{c}{$M_n$} &
$n$ & \multicolumn{1}{c}{$M_n$} & 
$n$ & \multicolumn{1}{c}{$M_n$} \\ \hline 
1& 1 &     9  & 933458 &         17 & 59923200729046  \\                 
2& 2 &     10 & 8152860 &        18 & 608188709574124 \\                  
3& 8 &     11 & 73424650 &       19 & 6234277838531806 \\                 
4& 42&     12 & 678390116 &      20 & 64477712119584604 \\                 
5& 262&    13 & 6405031050 &     21 & 672265814872772972 \\              
6& 1828 &  14 & 61606881612 &    22 & 7060941974458061392 \\              
7& 13820 & 15 & 602188541928 &   23 & 74661728661167809752 \\              
8& 110954 &16 & 5969806669034 &  24 & 794337831754564188184 \\            
\hline \hline
\end{tabular}
\end{center}
\end{table}

The enumerations undertaken thus far are too numerous to detail here. We 
thus only give the results for $M_n$ which are listed in 
Table~\ref{tab:closed}.  The series for the meander generating function
is characterised by coefficients which grow exponentially, with sub-dominant 
term given by a critical exponent. The generating function behaviour is 
$M(x) =\sum_n M_n x^n \sim A(x)(x_c - x)^{\xi},$ and hence the 
coefficients of the generating function 
$M_n = [x^n]M(x) \sim  \sigma^n/n^{\xi+1}\sum_i c_i/n^{f(i)} $,
where $\sigma=1/x_c$ is the connective constant. We analyzed the series  
by the numerical method of differential approximants \cite{Guttmann89},
and obtained the very accurate estimates $x_c=0.08154695(10)$ and
$\xi =2.4206(4)$, and we thus find that the connective constant
$\sigma=12.262874(15)$. Having obtained these accurate estimates we
used them to fit the asymptotic form of the coefficient to the
formula above. The results were fully consistent with $f(i) =i$.
There were no signs of half-integer or other powers, showing
that there does not appear to be any non-analytic correction terms to the
generating function. 

\section{Conclusion\label{sec:conclusion}}

We have presented an improved algorithm for the enumeration of closed
meanders. The computational complexity of the algorithm is estimated to 
be $2.5^n$, much better than direct counting algorithms which have 
complexity $12.26^n$. Implementing this algorithm enabled us to obtain 
closed meanders up to order 24.  From our extended series we obtained 
precise estimates for the connective constant and critical exponent.
An alternative analysis provides very strong evidence for the absence
of any non-analytic correction terms to the proposed asymptotic form
for the generating function.

\section*{Acknowledgments}

This work was supported by a grant from the Australian Research Council.

\end{document}